\documentstyle[epsfig]{aipproc}

\begin{document}
\def\lsim{{\buildrel < \over\sim}}
\def\gsim{{\buildrel > \over\sim}}
\def\to{\rightarrow}
\def\fb{~{\rm fb}}
\def\pb{~{\rm pb}}
\def\ev{\,{\rm eV}}
\def\kev{\,{\rm KeV}}
\def\mev{\,{\rm MeV}}
\def\gev{\,{\rm GeV}}
\def\tev{\,{\rm TeV}}
\def\wh{\widehat}
\def\wt{\widetilde}
\def\mhalf{m_{1/2}}
\def\gl{\wt g}
\def\q{$q$}
\def\qbar{$\bar{q}$}
\def\g{$g$}
\def\dc{$\delta_c$}
\def\als{\alpha_s}
\newcommand{\as}{{\ifmmode \alpha_S \else $\alpha_S$ \fi}}
\def\ttbs{\char'134}
\def\AmS{{\protect\the\textfont2
  A\kern-.1667em\lower.5ex\hbox{M}\kern-.125emS}}
\catcode`@=11
\def\Biggg#1{\hbox{$\left#1\vbox to 22.5\p@{}\right.\n@space$}}
\catcode`@=12
\newcommand\refq[1]{$^{#1}$}
\newcommand\ind[1]{_{\rm #1}}
\newcommand\aopi{\frac{\as}{\pi}}
\newcommand\oot{\frac{1}{2}}
\newcommand\sinsthw{\sin^2(\theta_{\rm W})}
\newcommand\logmu{\log(\mu^2/\Lambda^2)}
\newcommand\Lfb{\Lambda^{(5)}}
\newcommand\Lfc{\Lambda^{(4)}}
\newcommand\Lf{\Lambda_5}
\newcommand\epem{\ifmmode e^+e^- \else $e^+e^-$ \fi}
\newcommand\mupmum{ \mu^+\mu^- }
\newcommand\bbar{b\bar{b}}
\newcommand\gamgam{\gamma\gamma }
\newcommand\ms{\ifmmode{\overline{\rm MS}}\else $\overline{\rm MS}$\ \fi}
\newcommand\Q[1]{_{\rm #1}}
\newcommand\pplus[1]{\left[\frac{1}{#1}\right]_+}
\newcommand\plog[1]{\left[\frac{\log(#1)}{#1}\right]_+}
\newcommand\sh{\hat{s}}
\newcommand\epbar{\overline\epsilon}
\newcommand\nf{\alwaysmath{{n_{\rm f}}}}
\newcommand\MSB{\ifmmode{\overline{\rm MS}}\else $\overline{\rm MS}$\ \fi}
\newcommand{\aem}{\alpha_{\rm em}}
\newcommand{\nlf}{\alwaysmath{{n_{\rm lf}}}}
\newcommand{\ep}{\epsilon}
\newcommand{\aop}{\frac{\as}{2 \pi}}
\newcommand{\Tf}{{T_{\rm f}}}
\newcommand{\mub}{\ifmmode \mu{\rm b} \else $\mu{\rm b}$ \fi}
\newcommand\alwaysmath[1]{\ifmmode #1 \else $#1$ \fi}
\newcommand{\TeV}{{\rm TeV}}
\newcommand{\GeV}{{\rm GeV}}
\newcommand{\MeV}{{\rm MeV}}
\newcommand{\LQCD}{\ifmmode \Lambda_{\rm QCD} \else $\Lambda_{\rm QCD}$ \fi}
\newcommand{\LMSB}{\ifmmode \Lambda_{\overline{\rm MS}} \else
          $\Lambda_{\overline{\rm MS}}$ \fi}
\newcommand{\qb}{\overline{q}}
\def\pp{\ifmmode p\bar{p} \else $p\bar{p}$ \fi}
\def\VEV#1{\left\langle #1\right\rangle}
\def\LMSb{\ifmmode \Lambda_{\rm \overline{MS}} \else
$\Lambda_{\rm \overline{MS}}$ \fi}
\def\ie{\hbox{\it i.e.}{}}      \def\etc{\hbox{\it etc.}{}}
\def\eg{\hbox{\it e.g.}{}}      \def\cf{\hbox{\it cf.}{}}
\def\etal{\hbox{\it et al.}}
\def\dash{\hbox{---}}
\def\abs#1{\left| #1\right|}   
\def\to{\rightarrow}
\def\d{{\rm d}}
\newcommand{\lra}{\leftrightarrow}
\newcommand{\la}{\langle}
\newcommand{\dd}{{\rm d}}
\newcommand{\PS}{{\rm PS}}
\newcommand{\pperp}{p_{\perp}}
\newcommand{\ra}{\rangle}
\def\vspaceinarray{\nonumber ~&~&~\\}

\newcommand{\Nc}{N_c}
\newcommand{\Nf}{N_f}
\def\LO{leading order }
\newcommand{\eps}{\epsilon}
\newcommand{\ve}{\varepsilon}
\newcommand\epb{\overline{\epsilon}}
\newcommand{\be}{\begin{equation}}
\newcommand{\ee}{\end{equation}}
\newcommand{\bea}{\begin{eqnarray}}
\newcommand{\eea}{\end{eqnarray}}
\newcommand{\beas}{\begin{eqnarray*}}
\newcommand{\eeas}{\end{eqnarray*}}
\def\abs#1{\left| #1\right|}
\def\Am{{\cal A}}
\def\nn{\nonumber}
\def\phys{{\rm phys}}
\def\ms{$\overline{{\rm MS}}$}
\def\limes#1{\mathrel{\mathop{\lim}\limits_{#1}}}
\def\arrowlimit#1{\mathrel{\mathop{\longrightarrow}\limits_{#1}}}
\def\mus#1#2{\left(-\frac{\mu^2}{s_{#1#2}}\right)^\varepsilon}
\def\qb{\bar{q}}
\def\Qb{\bar{Q}}

\relax
\def\ap#1#2#3{
        {\it Ann. Phys. (NY) }{\bf #1} (19#3) #2}
\def\app#1#2#3{
        {\it Acta Phys. Pol. }{\bf #1} (19#3) #2}
\def\ar#1#2#3{
        {\it Ann. Rev. Nucl. Part. Sci. }{\bf #1} (19#3) #2}
\def\cmp#1#2#3{
        {\it Commun. Math. Phys. }{\bf #1} (19#3) #2}
\def\cpc#1#2#3{
        {\it Comput. Phys. Commun. }{\bf #1} (19#3) #2}
\def\ijmp#1#2#3{
        {\it Int .J. Mod. Phys. }{\bf #1} (19#3) #2}
\def\ibid#1#2#3{
        {\it ibid }{\bf #1} (19#3) #2}
\def\jmp#1#2#3{
        {\it J. Math. Phys. }{\bf #1} (19#3) #2}
\def\jetp#1#2#3{
        {\it JETP Sov. Phys. }{\bf #1} (19#3) #2}
\def\ib#1#2#3{
        {\it ibid. }{\bf #1} (19#3) #2}
\def\mpl#1#2#3{
        {\it Mod. Phys. Lett. }{\bf #1} (19#3) #2}
\def\nat#1#2#3{
        {\it Nature (London) }{\bf #1} (19#3) #2}
\def\np#1#2#3{
        {\it Nucl. Phys. }{\bf #1} (19#3) #2}
\def\npsup#1#2#3{
        {\it Nucl. Phys. Proc. Sup. }{\bf #1} (19#3) #2}
\def\pl#1#2#3{
        {\it Phys. Lett. }{\bf #1} (19#3) #2}
\def\pr#1#2#3{
        {\it Phys. Rev. }{\bf #1} (19#3) #2}
\def\prep#1#2#3{
        {\it Phys. Rep. }{\bf #1} (19#3) #2}
\def\prl#1#2#3{
        {\it Phys. Rev. Lett. }{\bf #1} (19#3) #2}
\def\physica#1#2#3{
        { Physica }{\bf #1} (19#3) #2}
\def\rmp#1#2#3{
        {\it Rev. Mod. Phys. }{\bf #1} (19#3) #2}
\def\sj#1#2#3{
        {\it Sov. J. Nucl. Phys. }{\bf #1} (19#3) #2}
\def\zp#1#2#3{
        {\it Zeit. Phys. }{\bf #1} (19#3) #2}
\def\tmf#1#2#3{
        {\it Theor. Math. Phys. }{\bf #1} (19#3) #2}

\begin{flushright}
{\large ETH--TH/97--27}\\
{\rm September 1997\hspace*{.5 truecm}}\\
\end{flushright}

\vspace*{3truecm}

\begin{center}
{\Large \bf 
The physics potential of the LHC 
\footnote{Talk given at the meeting
 Beyond The Standard Model V, Balholm, Norway
 April 29 -- May 4, 1997\,;
E-mail:
 kunszt@itp.phys.ethz.ch} }\\[0.5cm]
 {\large 
Zoltan ~Kunszt}\\[0.15 cm]
{\it  Institute of Theoretical Physics, ETH, Z\"urich, Switzerland.}\\[0.15cm]
\end{center}
\vspace*{2truecm}
\begin{abstract}
\noindent\small 
This talk is a short overview of the physics potential of the LHC 
with emphasis on  Higgs search and SUSY search.
First I review why 
LHC with the ATLAS and CMS detectors is expected  to give a decisive
test of the electroweak symmetry breaking mechanism of the Standard Model.
Then I consider  the Higgs sector of the
Standard Model (SM).
Finally the search for supersymmetry
is  discussed within the framework of 
 various implementation of the Minimal Supersymmetric Standard Model (SM).

\end{abstract}
\vspace*{2cm}

\section*{Introduction}
Future  progress concerning the physics beyond the
Standard Model  depends crucially from  the successful 
experimental tests of  the electroweak symmetry breaking
mechanism. The Standard Model is an
``almost good'' effective  field theory
with many good and few bad properties\cite{efffield}.
According to the effective field theory concept the Lagrangian
of a successful low energy theory is built 
from  the known fundamental fields
 restricted by the symmetries of the
theory. Renormalizability is not required, but it  emerges as
the consequence of the fact that the scale of new physics
is high in comparison with the experimentally accessible energy range.
The effective Lagrangian  then can be classified into
 relevant, marginal
and irrelevant terms
\be
{\cal L}_{\rm eff}= {\cal L}_{\rm relevant}+ {\cal L}_{\rm marginal}+
{\cal L}_{\rm irrelevant} 
\ee
or in terms of local operators
\be
{\cal L}_{\rm eff}=\sum_{n,i}\frac{c_n^{(i)}}{\Lambda^{n-4}}{\cal O},\qquad
 {\rm dim}_{\rm mass} {\cal O}_n=n
\ee
The relevant, marginal and irrelevant terms are those with
$n<4$ (fermionic and bosonic mass terms), $n=4$ (kinetic
energy terms, gauge interactions, Yukawa interactions, Higgs
bosons self interaction) and $n>4$ ( non-renormalizable interaction
terms such as $(\overline{\psi}\psi)^2$). The contributions
of irrelevant terms are negligible at low energies since they 
give negligible terms of order $\approx (E/\Lambda)^{n-4}$ inversely
proportional to some positive power of the cut-off. In contrary,
the relevant terms
 give rise to  contributions proportional to the cut-off,
and   are unwanted since
they can not hide efficiently the effects of the unknown physics of
very high scales.
Therefore a {\it good} low energy
effective theory is a renormalizable theory with  only marginal terms
\be
{\cal L}_{\rm eff}= \sum {\rm all\  marginal\  operators}
\ee
The Standard Model because of 
two  (apparently minor) faults is NOT 
a good effective field theory.
First, one
relevant term is missing.
According to the general rules of constructing good effective 
field theory in QCD we have to include CP violating the marginal term
\be
{\cal L}_{\rm QCD}^{(\Theta)}\approx \Theta
\epsilon^{\mu\nu\lambda\sigma}
G_{\mu\nu}G_{\lambda\sigma}
\ee
where $G_{\mu\nu}$ denotes the gluon field tensor.
The experimental limit on the coupling of this   term, however, is 
extremely small $\Theta< 10^{-9}$. This is puzzling and lead to
the suggestion of the existing of axions.
The second fault is an unwanted relevant term.
Although
local gauge invariance forbids   mass terms for vector
particles and the requirement of  chiral  $SU(2)_L\times U(1)$ gauge symmetry  
for right handed fermion singlets and
left handed fermion doublets forbids fermionic mass terms, 
a scalar mass term of  the Higgs-boson,   a
relevant operator, is allowed.
 The presence of a relevant
scalar mass term of the SM Lagrangian gives the most clear hint
that its range of the validity  can not extend to 
far above the scale of the electroweak symmetry breaking of $\approx
260 \GeV$. 
This  qualitative argument is independent from the actual 
value of the Higgs mass and this difficulty    (in slightly different context)
 is called the gauge
hierarchy problem. It  may indicate that either the
electroweak
symmetry breaking mechanism is not given simply by elementary Higgs
fields or that there is a new additional  symmetry which renders the
relevant operators  of the scalar sector to marginal.
 Technicolour
models give  examples for the first possibility and  supersymmetry
with low  supersymmetry breaking scale is the answer for 
the second possibility. 
For example, 
  the Minimal Supersymmetric Standard Model is favored
in comparison with the SM since
 it provides a ${\it good}$ effective field theory.  

During the first year LHC will operate 
as proton-proton collider at center-of-mass energy of 
$\sqrt{s}=14\tev$ with low luminosity 
 ${\cal L}=10^{33}$ ${\rm cm}^{-2}$ ${\rm sec}^{-1}$
which subsequently will be increased to the design value
of ${\cal L}=10^{34}$ ${\rm cm}^{-2}$ ${\rm sec}^{-1}$.
 With the universal   ATLAS\cite{ATLAS} and CMS\cite{CMS} detectors 
in its proton-proton collider mode LHC will provide us the 
possibility to  test the Standard Model well above the 
scale of the electroweak symmetry breaking\cite{aachenpr}.
 It will be possible 
to test the validity of the one-doublet Higgs sector of the SM as
well its possible supersymmetric extensions. One can decisively test
the MSSM standard model and  get direct
evidence for the existence of the 
supersymmetric
partners of the known particles.
If elementary Higgs boson does not exist the ATLAS and CMS
collaborations  will  be able to  provide
the first experimental hint  for the presence of
  new type interactions between longitudinal W-bosons.
Beyond these fundamental physics tests, 
 new quarks or leptons, new electroweak gauge
bosons or leptoquarks could be discovered or 
the decay mode of the top quarks could be quantitatively tested.
LHC will also be able to accelerate heavy ions and for example
by observing $Pb-Pb$ collisions at 1150 \tev\, center of mass energy at
luminosity of $10^{27}$ ${\rm cm}^{-2}$ ${\rm sec}^{-1}$  
with the ALICE detector 
 it will be possible to obtain decisive  experimental test
on the  physics of strongly interacting matter at extreme energy
densities. In particular it will be possible to test
 the formation of  quark-gluon
plasma, a new phase of matter. Finally, special purpose detector will be
installed to perform high precision experiment on B-physics, with emphasis
on CP-violation.

The full use of the physics potential requires extreme effort in
the performance of the machine and the detectors.
In order to illustrate the new technical
complications at LHC I recall that in the pp collider mode as a result
of  the very
large non-diffractive inelastic cross-section of about $70$ mb
 on average 18 minimum bias interactions are expected
per beam crossing in 25 ns time intervals.
The interesting  weak physics signals of short distance
physics  are buried in this
enormously noisy background.
The weak signatures of new physics can show up  in a number 
 of (sometimes complex) final states of leptons, jets and missing energy. 
This puts extreme requirement on the performance of the 
detectors : they must have  good particle identification, good
energy, momentum and angle resolutions for charged leptons, jets,
photons and missing transverse energy. ATLAS and CMS are designed to
meet these constraints.

\section*{Search for the Higgs Boson of the Standard Model}
The search method for the SM Higgs boson depends crucially on
the actual value of its  mass $m_H$\cite{guide}, therefore,   constraints 
restricting the allowed values of $m_H$ have great significance.
The latest analysis  of the LEP experiments\cite{lephiggs}   gives 
experimental lower bound $  m_H > 77 \gev$ at 95 \% confidence level 
as a result of direct search.
The final analysis at the end of  the LEP200 program
is expected to give a lower limit of about $95-100\gev$.
The high precision data are consistent with the theoretical
predictions (which depend on  $m_H$ via higher order radiative
corrections) if $m_H<430\gev$ at 95\% confidence level.

There are, however,  also important theoretical constraints. The triviality
problem constraints the range of
 validity of the field theoretical treatment of the scalar
sector  characterized by a cut-off scale. This cut-off 
scale can be considered as the upper limit on the scale of  new physics.
It is, however,   correlated to the value of the Higgs
mass. In   perturbation theory, in leading order, 
 the running coupling of the quartic scalar
self-interaction $\lambda(\mu)$ has a Landau pole at $\Lambda_c$
\be\label{mhiglam}
\lambda(\mu)=\frac{\lambda(m_H)}
{1 - 12\frac{\lambda(m_H)}{16\pi^2}\ln \frac{\mu^2}{m_H^2}}
,\qquad \Lambda_c=m_H e^{\frac{2\pi^2}{3\lambda(m_H)}}
\ee
such that $\lambda(\Lambda_c)=\infty$, where  in leading
order $\lambda(m_H)=m_H^2/2v^2$ and $v=256\gev$.
At small Higgs mass the cut-off can be as large as the Planck mass
$10^{19}\gev$ and it decreases exponentially with increasing Higgs mass.
By demanding  the cut-off scale to be larger
than the Higgs mass we get
 an upper bound on $m_H$.
This  perturbative result are
 further substantiated by non-perturbative treatment.
 L\"uscher and Weiss\cite{luescherweiss}
 found that at the scale where the sensitivity to 
the cut-off becomes non-negligible the value of the Higgs mass
is about $650 \gev$, moreover,  the value of 
$\lambda(2M_H)$ is about 3.5 well within 
 the perturbative regime.
It has  recently been shown\cite{riesselmann} that this  result is consistent
with  perturbation theory if higher order  perturbative corrections
( at least 
 two loop order) are taken into account.
The two loop beta function 
\be
\beta_{\lambda}=24\frac{\lambda^2}{(16\pi^2)^2}-312
\frac{\lambda^3}{(16\pi^2)^3}
\ee
develops a metastable fixed point at  $\lambda_{\rm FP}\approx 12.1$.
A typical  cut-off scale can be defined by requiring
$\lambda(\Lambda_c) \approx \lambda_{\rm FP}/2$.
The upper limit obtained from such an analysis is consistent with the
lattice values, furthermore one also finds that
$ W_L W_L $ scattering is well described by improved NLO
perturbation theory up to $\sqrt{s}=2\tev$ collision energies.
This result means that in the SM we can  describe  $W_LW_L$ 
scattering at LHC and NLC precisely.
\begin{table}[t!]
\caption{Upper and lower limits on the SM Higgs mass
at typical values of the cut-off.} 
\label{table1}
\begin{tabular}{|c|c|c|}
cut-off in (\gev)& upper limit (\gev)&lower limit (\gev) \\
\hline
$10^3$ & $650\pm 150$ &$ 47\pm 4$   \\
$10^6$ & $300\pm 20$ & $120\pm 8$   \\
$10^{15}$ &$ 195\pm 5$ & $140\pm 10$   \\
$10^{19}$ &$ 180\pm 4$ & $147\pm 12$   \\
\end{tabular}
\end{table}

For quantitative studies, in the the evolution of $\lambda(\mu)$ the large
Yukawa coupling of the top quark has to be taken into account.
\be 
\beta_{\lambda}=(24\lambda^2+12\lambda g_t^2-6g_t^4)/(16\pi^2)^2\, +\, 
{\rm\  gauge\ and \ higher \ order \ terms}
\ee
where in leading order $g_t^2(m_t)=\sqrt{2}m_t/v$. Since the top quark
is heavy
$m_t=175\gev$ these new terms  can  drive $\lambda(\mu)$  to negative
values at small scales.  By demanding the stability of the vacuum
($\lambda>0$) we get
a lower limit
on the Higgs mass.  In recent two loop calculations\cite{altisi} (
when the QCD corrections are non-negligible) the lower bounds have
been evaluated with  $m_t=175\pm 5\gev$ and $\as=0.118\pm.003$. In Table 1
we summarized the upper and lower bounds for a few typical value of
the cut-off\cite{hambriess}.
The LEP limits on $m_H$ quoted above are completely consistent with the
theoretically allowed range, in particular, the values $m_H \approx 170\gev$ 
allow a valid field theoretical treatment of the SM model up to the
Planck scale. In this case the bound following from  triviality alone
is much higher than the scale of new physics  suggested by the 
gauge hierarchy problem.
These considerations as well as the first indirect experimental hints
 indicate  that the mass of the Higgs boson is likely in 
the mass range of $m_H=95-400\gev$. 
Particularly difficult of the detection of the Higgs boson in the
low mass range  $m_H=95-140\gev$.
\begin{figure}[t!] 
\centerline{\epsfig{file=
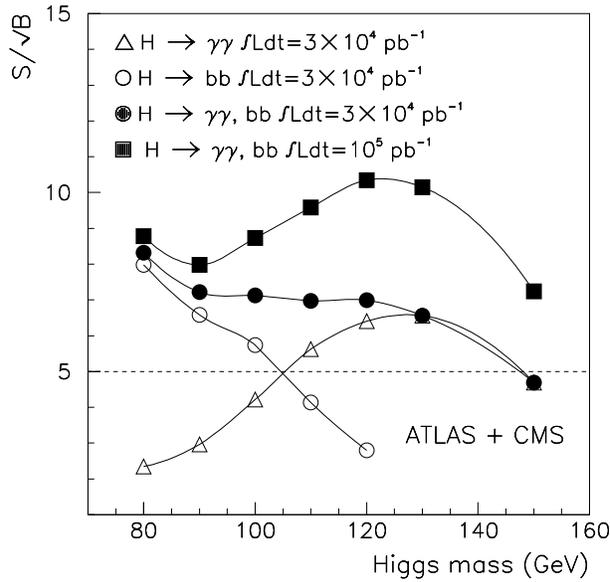,height=3.5in,width=3.5in}}
\vspace{10pt}
\caption{Ratios of signal over square root of  background rates for various
SM Higgs-production mechanisms as a function of the Higgs mass.}
\label{fig1}
\vspace*{-.2truecm}
\end{figure}
The signal to 
background ratio plotted  in fig.~\ref{fig1}~\cite{richter} clearly shows that the
full coverage of this range requires a run at high luminosity, the
combination of the signals of several  decay modes ( $h\to \bbar$ and
$h\to \gamgam$) and detector performances as designed or better. 
 In the the intermediate  
range $m_H=140-180\gev$ the detectable signals  are provided  by
the four lepton decay modes  of  gauge boson pair ($ZZ^*$  and $WW^*$)
production.
 Recently Dittmar and
Dreiner~\cite{dittdrei} have shown that 
the  polarization properties of the $WW$-pairs if come from  Higgs decay
are very different from the ones of those $WW$-pairs which are produced 
by the standard QCD quark-antiquark annihilation mechanism.
Therefore, in the range $m_H=155-180\GeV$  even with  
two undetectable neutrinos in the final state
 the $WW^*$ pair production and its subsequent
decay into $l^+l^{'-}\nu_l\overline{\nu}_{l'}$ gives improved
  signal together with the
  mode $H\to Z^0Z^{0*}\to l^+l^-l^+l^-$.
 Finally in the upper range $m_H=180-700\gev$ the gold plated
mode $H\to Z^0Z^{0}\to l^+l^-l^+l^-$ gives clear signal.
Updated branching ratio and cross-section values
as well as the discussion of the background can be found in 
refs.~\cite{MWJSZK,spirahab,JGSWS}.

\section*{Search for Supersymmetry}
 As was mentioned in the introduction supersymmetry offers a solution to
 the gauge hierarchy problem.
With   direct  supersymmetrization 
of the SM we get a better effective field theory provided the scale
of supersymmetry breaking is naturally low (around $1\tev$).
Supersymmetry is a theoretically attractive concept
since it gives  a generalization of Poincare invariance and 
it emerges naturally in string theories.
\subsection*{MSSM with SUGRA universal soft breaking terms}
The MSSM is defined by  minimal direct supersymmetrization
of the Standard Model with supersymmetric GUT and with
some simple boundary condition at the GUT scale.
It has four basic properties.
(1)
  it has minimal gauge group: $SU(3)_C\times 
SU(2)_L\times U(1)_Y$, (2) it 
has minimal particle content: three generation of quarks and leptons and
 their super partners, and two Higgs doublets plus superpartners;
 (3)
 it has an exact discrete $R$-parity, with $R=+1$ for SM particles and
Higgs-bosons and $R=-1$ for the superpartners;
 (4) its couplings are constrained
by  SU(5) GUT
with  universal soft
breaking terms at the GUT scale.
Contrary to the SM, the SU(5) GUT MSSM is consistent with limit on the proton 
decay. 

The assumption of SU(5) GUT unification with universal soft breaking 
allows to  reduce the huge number of soft breaking terms.
This  universality is  motivated by
 a universal supergravity Higgs mechanism in a   hidden sector.
Aside from the SM parameters, the model is completely specified by
five SUSY breaking parameters: $m_0$ (universal scalar mass), 
$m_{1/2}$ (universal gaugino mass) , bilinear and trilinear scalar couplings
$\mu,A_0,B_0$. A very nice feature of the model is that it gives rise to 
radiative  electroweak symmetry breaking in a wide range of parameters.
Because of R-parity, the lightest SUSY particle (LSP) is stable which leads 
to the important missing transverse energy signal.
 The phenomenology of the SUGRA MSSM is well known with experimentally
allowed constraints on its parameters.
At the LHC the full range of possible supersymmetric particle masses can be
explored.
Extensive studies determined the regions of parameter space for direct discovery.
The typical physics signals are i) $l^{\pm}+ $ jets $+$ missing$\, E_T$
for  gluino and squark pair production up to mass values of $3.6\tev$
ii) $\l^{\pm}l^{\pm}$ $+$   jets $+$ missing $ E_T$
for  $\tilde{g}\tilde{g}, \tilde{g}\tilde{q}.\tilde{q}\tilde{q}$
production. iii) $\l^{\pm}l^{\pm}l^{\mp}+ $ missing $E_T$ for
$\tilde{\chi}_1^0\tilde{\chi}_1^{\pm}$ production 
and  iv) $l^{\pm}l^{\mp}+$ missing $E_T$ for  slepton pair production
up to $300\gev$ sleptons. If   the superpartners
 will be found  their parameters can be measured 
by good  precision.
 SUGRA MSSM gives a good strategy for searching for 
certain type of signals of supersymmetry, 
 however,  it has several specific features
which are strongly depend on its untested technical simplifying assumptions.
Therefore it is important to consider some other viable MSSM models
with different soft breaking terms giving different supersymmetric
particle spectrum and so different physics signals. It is also
interesting to study R-parity violating schemes in which case the production 
of supersymmetric particles in general will not give the celebrated 
missing $E_T$ signals.
\subsection*{MSSM with gauge sector mediated SUSY breaking}
 Recently,  motivated by the
anomalous CDF $ee\gamgam$ event the phenomenology
of the MSSM with  gauge mediated SUSY 
breaking~\cite{dinenelson,pierce} also have been considered in great detail.
 In this case
the lightest
supersymmetric particle is the light gravitino therefore the 
lightest neutralino can decay as $\chi_1^0\to \tilde{G}+\gamma$.
Cosmological constraints give the upper limit $m_G<1\,\kev$.
The parameter space of this model is strongly constrained by  LEP200 where
a large part of it can    be excluded in the case of smaller values
of the SUSY breaking scale when the next-to-lightest supersymmetric
particle (NLP) can decay in the detector.
Assuming higher SUSY breaking scale
 the NLP will decay 
outside the LHC detector which may require 
the building of dedicated detectors\cite{maki}.

\subsection*{MSSM with R-parity violation}
Recently, the ALEPH anomalous four jet events and the anomalous HERA events
called great attention to the  detailed
  phenomenology of  MSSM models
 with $R$-parity violation\cite{Rpardreiner}. $R$-parity conservation is not a unique mechanism
to prevent fast proton decay.
Two less constraining mechanisms are the 
so called  baryon or lepton parities. The most general superpotential
has  both  soft and hard $R$ parity violating terms
\be
-{\cal L}^{\rm soft}_R=\mu_iLL_i H^1
\ee
\be
{\cal L}^{\rm hard}_R = \lambda_{ijk} L_i L_j E^c_k+ 
\lambda^{'}_{ijk} L_i Q_j D^c_l+ \lambda^{''}_{ijk} U^c_i D^c_j D^c_k
\ee
where $L$ denotes the chiral superfield of the lepton doublet, $U^c$
is the anti u-quark singlet and $Q$ is  the quark doublet.

The first two terms give rise 
 lepton number violating transitions  $\Delta L=1$
while the third term gives baryon number violating transitions $\Delta B=-1$.
The transition amplitude of fast proton decay is proportional
to the product  $\lambda^{'} \lambda^{''}$.
$R$-parity is a $Z_2$ symmetry which forbids all the three R-parity 
violating couplings.  One can, however,  impose  weaker
discrete $Z_3$ symmetries. In the the case of
$Z_3$ baryon parity we get  $\lambda ^{''}=0$ and in the case of 
$Z_3$ lepton parity one gets $\lambda=\lambda^{'}=0$.
In both case fast proton decay is forbidden.
Since we do not understand the origin of these discrete symmetries
 all the three options have  to be 
considered.
The unattractive feature  of the $R$-parity violating options is
that there are 45 additional  couplings. Their values 
are only constrained by low energy data.
  The most important property of  $R-$parity
violation is that the celebrated missing energy signal  got lost.
For example, a selectron can decay with $R-$ parity violating coupling
into two jets or a stop quark can decay into a $d$-quark and a positron
when the stop production would give 
 signals similar to the production of a leptoquark.
The phenomenology of the MSSM with R-parity violating terms
is not yet worked out at the necessary details but work is in progress
in this direction.
\begin{figure}[t!] 
\centerline{\epsfig{file=
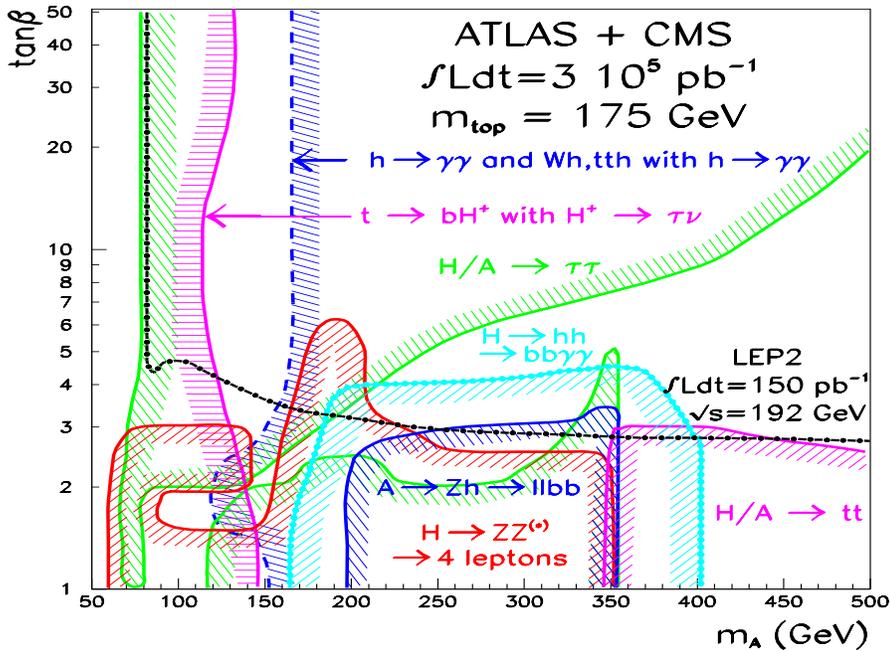,height=3.5in,width=3.5in}}
\vspace{10pt}
\caption{Regions of the parameter space $(\tan\beta - m_A)$
defined by $5\sigma$-discovery contours  for various MSSM
signals at high integrated luminosity of $3\times 10^5 \pb^{-1}$}
\label{fig2}
\end{figure}

\subsection*{Search for the MSSM Higgs Bosons}
The search for  the Higgs bosons of the MSSM has particular
significance in searching  for supersymmetry 
 since 
the Higgs sector of the MSSM is
largely  independent from the specific 
technical assumptions on the soft breaking terms and so from the 
properties of  the supersymmetric particles.

The physical states of the MSSM Higgs sector are  three neutral
bosons
(two CP-even, h and H, and one CP-odd, A) and a charged Higgs boson
$H^{\pm}$.
In the Born-approximation the MSSM Higgs sector contains only two
independent
parameters. A usual choice for these parameters 
are $m_A$, the physical mass of the CP-odd
neutral boson and  $\tan \beta=v_1/v_2$, where the vacuum 
expectation value $v_1$ gives mass to
the  quarks of charge $-1/3$ and to the leptons while $v_2$ gives
mass to the quarks of charge $2/3$. The parameter $m_A$ is essentially
unconstrained, although naturalness arguments suggest that it should
be smaller than ${\cal O}(500\gev)$ and $1<\tan\beta<m_t/m_b$.
There are important inequalities. The most important one is on the
mass of the light CP-even Higgs boson $m_h<M_Z \cos 2\beta$. This
leading order inequality is modified by radiative corrections
allowing  higher values up to $150\gev$,  therefore if the Higgs boson
will not be found at LEP200, the decisive test
should  come from the LHC. If $M_A$ is large than $200\gev$ or so  the
properties of the light Higgs boson $h$ is very similar to the SM
Higgs boson. 
 At low values of $\tan \beta$ the upper
limit is smaller.
 At the LHC 
the Higgs searches are  even more
difficult than the search for the SM Higgs since the production
rates are usually  smaller. There is 
 a variety of signatures  in which MSSM Higgs bosons can be
observed. Some of them similar to the SM case others concerns the
production of decay of the heavy CP-even  (H) and the CP-odd (A)
and charged Higgs bosons.
The search strategies and methods and the corresponding 
cross-section and branching ratio analysis have been carried out
long time ago\cite{zkfz}, a latest very detailed signal and background
analysis can be found in ref.~\cite{richter}.

 The discovery potential of LHC is summarized
in fig.~2\cite{richter} in terms of $5\sigma$ discovery contours in
($\tan~\beta\dash M_A$) parameter plane. The results are obtained
assuming no mixing in the third generation,
 $m_t=175\gev$ and a SUSY mass scale of $1\tev$.
Furthermore it was assumed 
that the decay modes of the Higgs bosons into supersymmetric
particles are unimportant. With comparing the results summarized on fig.~1 and
fig.~2 we can see that 
  the search for the MSSM Higgs bosons at LHC is in 
general more difficult than the search for the SM Higgs boson and  
 that the   constraints provided by LEP200  are crucially important
for a decisive test.

\end{document}